\documentclass[twocolumn,english]{elsart3p}
\usepackage[T1]{fontenc}
\usepackage[latin1]{inputenc}
\usepackage{float}
\usepackage{amsmath}
\usepackage{graphicx}
\usepackage{amssymb}

\makeatletter

\usepackage{float}
\usepackage{amsfonts}

\usepackage{babel}
\makeatother
\begin{document}
\begin{frontmatter}

\title{A new numerical algorithm for Low Mach number supercritical fluids}

\author{Jalil Ouazzani$^{1,2}$ and Yves Garrabos$^{1}$}

\address{$^{1}$\textbf{ICMCB, 87 Avenue Du Dr Albert Schweitzer, 33608 Pessac,
France} }

\address{$^{2}$\textbf{ArcoFluid, 4 All\'{e}e Du Doyen Georges Brus, 33600,
Pessac, France}}

\begin{abstract}
A new algorithm has been developed to compute low Mach Numbers supercritical
fluid flows. The algorithm is applied using a finite volume method
based on the SIMPLER algorithm. Its main advantages are to decrease
significantly the CPU time, and the possibility for supercritical
fluid flow modelisation to use other discretisation methods (such
as spectral methods and/or finite differences) and other algorithms
such as PISO or projection. It makes it possible to solve 3D problems
within reasonable CPU time even when considering complex equations
of state. The algorithm is given after first a brief description of
the previously existing algorithm to solve for supercritical fluids.
The side and bottom heated near critical carbon dioxide filled cavity
problems are respectively solved and compared to the previously obtained
results.
\end{abstract}
\end{frontmatter}

\section{Introduction}

In the last decade, numerous numerical works have been devoted to
the modeling of supercritical fluids (SCF)\cite{Onuki1990,Zappoli1990,Zappoli1996,Amiroudine2001,Accary2005,Polezhaev2005,Accary2006},
especially near their gas-liquid critical point. In fact, such fluids
exhibit quite unusual properties, behaving as highly expandable gases
with liquid-like density. An important point which has been pointed
out is the apparition of the so-called piston effect in a closed supercritical
fluid cell heated on a wall, where this piston effect acts like a
fourth mode of transport of energy. As a matter of fact, Onuki and
al \cite{Onuki1990} pointed out the thermodynamic importance of the
adiabatic heating while a more detailed hydrodynamic mechanism of
thermalization was proposed by Zappoli and al \cite{Zappoli1990}.
Close to the sample wall, heat diffusion makes a thin hot boundary
layer expand and compress adiabatically the rest of the fluid. A spatially
uniform heating of the bulk fluid occurs \cite{Onuki1990}with a thermalization
which should process at the velocity of sound \cite{Zappoli1990}.
Simultaneously, some dedicated experiments \cite{Guenoun1993,Garrabos1998,Frohlich2006}
have evidenced the existence of minute (a few $µm/s$, see \cite{Frohlich2006})
but really efficient flows at the border of the expanding diffusive
layer and compressed bulk fluid, independently of the closed geometries
under consideration \cite{Frohlich2006}. Such an effect is accounted
for into the transport equations through source terms, which are non-linearly
related between density, temperature and pressure, and a real equation
of state of supercritical fluid. 

The various research who have been conducted while treating critical
fluids have been done at very low mach numbers $M_{a}$ (where $M_{a}$=
flow velocity/speed of sound). Depending on the Mach number and on
the transient or not character of the flow, depending on the type
of fluid, ideal or very expandable as supercritical fluids are, the
relative strength of the coupling of mass, momentum, energy and equation
of state can be very disparate and one cannot use an universal efficient
method for all the range of Mach number. As a matter of facts the
dynamic coupling with pressure as introduced in high velocity flows
{}``propagates'' by the equation of state to the work of pressure
forces in the energy balance because of the strong pressure gradient.
In the case of very low velocity flows of ideal gazes this term is
vanishing and the energy balance is mainly governed by heat diffusion
while the momentum balance is that of a non compressible fluid in
the Boussinesq approximation. However when it came necessary to compute
unsteady flows of very expandable fluids like near critical fluids,
new problems arose. These fluids are characterized by diverging isothermal
expansion coefficient, isothermal compressibility, thermal conductivity
and heat capacity at constant pressure, their exponent being such
that the heat diffusion goes to zero when nearing the critical point
of the phase diagram. Their isentropic compressibility, linked to
the velocity of sound which goes to zero very slowly, remains comparable
to that of ideal fluids. This is why the term {}``hyper compressible
fluids'' as often encountered for supercritical fluid can be misleading
and it is better to say {}``hyper expandable fluids''. This mean
first that the coupling between velocity and pressure is still very
weak in the momentum balance equation. However due to the diverging
isothermal expansion coefficient, the equation of state is no longer
a passive link between density, pressure and temperature since it
makes the density variations in heat diffusion layer to be several
order of magnitude larger than the temperature ones. Accordingly,
the term of the energy balance which represents the work of the pressure
forces due to the deformation of a fluid element during its motion
becomes prominent, as it is in high velocity ideal fluid flows. The
consequence of this fact was the numerical difficulties encountered
during the first attempts to calculate heat propagation by thermoacoustic
coupling in supercritical fluids. In supercritical fluids where we
work in a closed geometries, the heat input generates a piston type
effect in which the isentropic compression due to the slow motion
generated by this piston effect is modified by acoustic waves. These
waves propagate back and forth many times, because they are reflected
at the walls of the domain. In that problem, we have one length scale,
let say domain size, and two times scales: the long time it takes
the slow diffusive flow to travel one length scale and the short time
it takes an acoustic waves to travel one length scale. Therefore,
the numerical approaches have opted for two solutions to completely
treat heat transfer phenomena: the first one considers the simulation
in time of the order of acoustic time and remains appropriate to account
for initial heating period, while the second one accounts for time
higher than the piston effect time up to diffusive time. In the first
case, the full transport equations are taken without any approximation,
whereas in the second case in order to have in the numerical algorithm
time steps which are not drastically small (in the order of acoustic
time), one has to resort to the low mach number filtering approximation
\cite{Zappoli1996}. The low mach number filtering decouples the density
from the dynamic pressure which is the pressure part dependent both
on space and time \cite{Paolucci1982,Muller1998,Munz2003}. The other
pressure part refers as a thermodynamic pressure in the following
text and it is a part dependent only in time, except when its accounts
for a possible local contribution due to the hydrostatic pressure
which plays a significant role in supercritical fluids under gravity
field, even for very small heights of the cell \cite{Accary2005}.
As a consequence the numerical algorithms chosen have been limited
to finite volume methods which require excessive amount of CPU time
due to the necessary iterative scheme for coupling all equations,
one has to keep in mind that due to the singularities arising near
the critical point direct methods of resolution have failed \cite{Accary2006}. 

In this paper we present an algorithm which decouples at each time
step the energy equation and the equation of state from the momentum
and mass conservation equations. This decoupling brings a substantial
reduction of the CPU time enabling a more straightforward extension
to 3D modeling. It also permits to use other methods than finite volumes,
i.e. finite differences and spectral methods, enables one to use besides
SIMPLE and SIMPLER algorithms\cite{Patankar1980}, more direct algorithms
such as PISO \cite{Oliveira2001} or a modified projection technique
\cite{Peyret1983}, and finally gives the possibility to extend simulation
using any complex equation of state (EOS). The algorithm is also applicable
to several situations encountered in low mach numbers flows as for
example in subsonic combustion problems, or, more generally, in flows
with velocities much smaller than speed of sound but having important
density variation with temperature.

The paper is organized as follows: Section 2 describes the previous
algorithm used in the modeling of SCF under the low Mach number approximation
and then describe the new algorithm. Section 3 shows numerical validation
for two typical examples, before concluding in Section 4.

\section{Numerical algorithm and Low mach number filtering}

In the present case, where we interest ourselves to critical fluids,
we have to point out first the major difficulties which arise when
approaching the critical point. Going to the limit close to the critical
point, the main physical properties are diverging, terms such as the
pressure work in the energy equation are becoming as in compressible
flows most leading terms and are in fact the terms in the equations
responsible for the {}``piston'' effect with a uniform rise of the
temperature in the bulk for closed domains\cite{Onuki1990}. In addition,
the most straightforward equation of state which we can use to describe
these real fluids is the van der Waals equation of state which is
non linear and for which a direct resolution is not possible close
to the critical point due to the fact that this equation degenerates
close to the critical point\cite{Accary2006}. Therefore one has to
use iterative method to solve for the density, the thermodynamic pressure
and the temperature. The convergence for this triplet $(P-\rho-T)$
becomes quite slow due to the above and due to the fact also that
in order to obtain the pressure work term in the energy equation,
we have to obtain the divergence of the velocity field through the
solving of the momentum equation and the equation of conservation
of mass.

Most of the interesting physics contained in studying the SCF are
inherently transient, this brings us to another aspect which is: even
tough we are in an hyper expandable fluid, the fluid velocities are
much smaller than the speed of sound for the cases we would like to
consider, i.e. small temperature perturbations of such fluids. Therefore
another difficulty appears which this time puts the stringed on the
size of the time step, the fluids being considered as low Mach number
fluids\cite{Paolucci1982,Muller1998,Munz2003}. We then encounter
the same difficulties as in low Mach number combustion, where the
acoustic pressure waves force the algorithm for not becoming unstable
to adopt small time steps: these time steps have to satisfy the CFL
condition that requires a time step size smaller than the grid size
times the reciprocal of the largest wave speed: \begin{equation}
\Delta t\leqslant\frac{\Delta x}{\max(c+\left|v\right|)}\label{eq1}\end{equation}
 where $c$ is the speed of sound, $v$ the flow velocity and $\Delta x$
the grid size.

There are two main approaches for solving low Mach number flows; the
first one, is to use compressible solvers (density based)\cite{Guillard1999};
and the second one, is in extending incompressible solvers (pressure
based) towards this regime\cite{Karki1989}. Both of these techniques
will anyway suffer from the pressure acoustic waves, in order to alleviate
these restrictions on time step two distinct techniques have been
proposed, preconditioning and asymptotic\cite{Guillard1999,Turkel1987,Choi1993,Chorin1997}.
Preconditioning techniques have many drawbacks for transient low Mach
number flows and therefore have not been considered in this work.
In the asymptotic technique or perturbation approach\cite{Muller1998},
a filtered form of the equations is employed to eliminate system stiffness.
We expand all the variables in Taylor series in power terms of the
Mach number, to do that we assume that the low Mach number asymptotic
analysis is a regular perturbation problem, i.e. all flow variables
can be expanded in power series of $M_{a}$ (flow velocity/speed of
sound) as for example the pressure which can be written as follow:
\begin{equation}
P(X,t,M_{a})=P_{0}(X,t)+M_{a}p_{1}(X,t)\label{eq2}\end{equation}
\[
+M_{a}^{2}p_{2}\left(X,t\right)+O\left(M_{a}^{3}\right)\]
 $P_{0}$, $p_{1}$ and $p_{2}$ are called the zeroth- (or leading),
first- and second-order pressure respectively.

The asymptotic analysis of the momentum equation implies that \begin{equation}
\nabla P_{0}=0\label{eq3}\end{equation}
 \begin{equation}
\nabla p_{1}=0\label{eq4}\end{equation}
 and it shows that for the purpose of solving the filtered transport
equations it is necessary only to retain the second order term in
the above expansion (\ref{eq2}) which yield to: \begin{equation}
P(X,t,M_{a})=P_{0}(t)+M_{a}^{2}p_{2}(X,t)+O(M_{a}^{3})\label{eq5}\end{equation}
 Equations (\ref{eq3}) and (\ref{eq4}) express that the zeroth-and
first order pressure terms in the series expansion of the total pressure
are independent on space and only dependent on time.

The other variables which are T the temperature, $\rho$ the density
and $\vec{V}$ the vector velocity with components U$_{i}$, are expanded
in the same manner in terms of $M_{a}$ number: \begin{equation}
\begin{array}{cl}
\vec{V}(X,t,M_{a})=\vec{V}_{0}(X,t) & +M\vec{_{a}V}_{1}(X,t)\\
 & +M_{a}^{2}\vec{V}_{2}(X,t)\\
 & +O(M_{a}^{3})\end{array}\label{eq6}\end{equation}
 \begin{equation}
\begin{array}{cl}
T(X,t,M_{a})=T_{0}(X,t) & +M_{a}T_{1}(X,t)\\
 & +M_{a}^{2}T_{2}(X,t)\\
 & +O(M_{a}^{3})\end{array}\label{eq7}\end{equation}
 \begin{equation}
\begin{array}{cl}
\rho(X,t,M_{a})=\rho_{0}(X,t) & +M\rho_{a1}(X,t)\\
 & +M_{a}^{2}\rho_{2}(X,t)\\
 & +O(M_{a}^{3})\end{array}\label{eq8}\end{equation}
 Again the asymptotic analysis shows that only the zeroth order terms
have to be retained in equations (\ref{eq6}-\ref{eq8}).

Using the above asymptotic expansions and replacing them in the transport
equations of real fluids with van der Waals equation of state, we
obtain the following dimensional equations (where $\vec{g}$ is the
gravity vector ):

i) Mass conservation equation: \begin{equation}
\frac{\partial\rho}{\partial t}+\frac{\partial\rho U_{i}}{\partial x_{i}}=0\label{eq9}\end{equation}

ii) Momentum equation: \begin{equation}
\begin{array}{cl}
\frac{\partial\rho U_{i}}{\partial t}+\frac{\partial\rho U_{j}U_{i}}{\partial x_{j}}= & -\frac{\partial p_{2}}{\partial x_{i}}\\
 & +\frac{\partial\left[\mu\left(\frac{\partial U_{i}}{\partial x_{j}}+\frac{\partial U_{j}}{\partial x_{i}}\right)\right]}{\partial x_{j}}\\
 & -\frac{2}{3}\mu\frac{\partial\left(\frac{\partial U_{i}}{\partial x_{j}}\right)}{\partial x_{i}}+\rho\vec{g}\end{array}\label{eq10}\end{equation}

iii) Energy equation (written here in term of $C_{v}$ and a viscous
dissipation term $\phi$, see below): \begin{equation}
\begin{array}{cl}
\frac{\partial\rho C_{v}T}{\partial t}+\frac{\partial\rho U_{j}C_{v}T}{\partial x_{i}}= & \frac{\partial\left[\lambda\left(T\right)\left(\frac{\partial T}{\partial x_{j}}\right)\right]}{\partial x_{j}}\\
 & -(P_{0}+a\rho^{2})\frac{\partial U_{i}}{\partial x_{i}}\\
 & +\phi\end{array}\label{eq11}\end{equation}

iv) The van der Waals equation of state (with $a$ and $b$ as fluid-dependent
parameters): \begin{equation}
P_{0}+a\rho^{2}=\frac{\rho T}{1-b\rho}\label{eq12}\end{equation}

The van der Waals equation leads to consider the isochoric specific
heat $C_{V}$ to be constant. Jointly to the use of this above classical
equation of state, the singular behavior of the thermal conductivity
$\lambda$ (estimated at constant critical density) is approximated
under the mean-field theory as follows:

\begin{equation}
\lambda\left(T\right)=\lambda_{b}\left(T\right)+\lambda^{MF}\tau^{-1/2}\label{eq:lamda}\end{equation}
where the label MF recalls for the mean-field value $\frac{1}{2}$
for the critical exponent of the power law in terms of $\tau=\frac{T-T_{c}}{T_{c}}$
{[}$T$ ($T_{c}$) is the (critical) temperature]. $\lambda_{b}$
is a background estimated far away from the critical point. The viscosity
coefficient $\mu$ (estimated at constant critical density) is approximated
by its background contribution in the mean field approximation. Therefore,
the viscous dissipation is written as: \[
\begin{array}{cl}
\phi= & 2\mu\left[\left(\frac{\partial U_{i}}{\partial x_{i}}\right)^{2}\right]+\mu\left[\left(\frac{\partial U_{i}}{\partial x_{j}}+\frac{\partial U_{j}}{\partial x_{i}}\right)^{2}\right]\\
 & -\frac{2}{3}\mu\left[\left(\frac{\partial U_{i}}{\partial x_{i}}+\frac{\partial U_{j}}{\partial x_{j}}\right)^{2}\right]\end{array}\]

More generally, we note that the van der Waals equation for thermodynamic
properties as mean field approximations for transport properties does
not only give us a phenomenological singular behavior for unusual
properties such as compressibility, heat capacity at constant pressure,
etc. but also needs to introduce analytical singular expression for
the transport properties as for example for the thermal conductivity
\eqref{eq:lamda}. However, we have considered the isochoric specific
heat at constant volume $C_{V}$ to be constant, as well as the viscosity
$\mu$, because, for both of these two properties, their values are
deduced from background contribution due to the fact that their critical
exponents of divergence are very low and that they can be neglected
for the values of distances to the critical point we are considering
in our numerical work ($\tau>10^{-4}$). Even tough the speed of sound
is decreasing when coming close to the critical point, it stays however
finite with values not lower than $50\, ms^{-1}$, implying that the
Mach number stays of the order of $10^{-5}$. Therefore under the
approximation of low Mach number where the filtering of acoustic waves
means that the time step in the numerical algorithm is bound only
by the flow velocity as opposed to the speed of sound, the above supercritical
fluid equations have been modeled exclusively with finite volume iterative
methods\cite{Patankar1980}. The choice for such iterative methods
was quite natural due to the fact that the main physical properties
have strong temperature variations, and that the divergence term in
the energy equation was the main term in exhibiting the so-called
{}``piston'' effect. In supercritical fluids, we cannot avoid the
iterative process coupling the thermodynamic pressure, the temperature
and the density. This has implied that other numerical methods such
as spectral methods\cite{Peyret1983}, finite differences and schemes
such as PISO or projection methods have been disregarded. In this
paper, we will start from the previously existing numerical algorithm
\cite{Zappoli1996}described below to introduce a new algorithm which
takes fully in account the low Mach number nature of the flow in closed
cavities for supercritical fluids, and through an Adams-Bashforth
scheme \cite{Peyret1983,Ouazzani1986,Nicou2000,Lessani2006,Frohlich1992}
enables us to avoid the iterative process for the momentum and continuity
equation leading to a substantial saving in CPU times.

\subsection{Description of algorithm 1}

We have first developed Cartesian and Polar three dimensional codes
where the discretisation of equations (\ref{eq9}-\ref{eq12})  used
a finite volume technique \cite{Patankar1980}leading to a finite
set of algebraic equations that can be solved iteratively in a segregated
manner. The solver chosen to accomplish such task are the BICGSTAB
for temperature and velocities, and a preconditioned conjugate gradient
for pressure. To couple momentum equation to continuity equation several
methods are available, but when turning to finite volume, the most
common methods used are the SIMPLE family type and its derivatives
(SIMPLEST, SIMPLEC, SIMPLER, PISO). The temporal scheme for convective
part has several choices which are: hybrid scheme, power law scheme,
Quick family schemes. For the examples shown later in this paper,
they have been treated using the power law scheme. The SCF equations
are inherently transient and imply to choose an algorithm which will
optimize the resolution at each time step. The ideal candidate should
be a modified PISO algorithm (a predictor -- corrector algorithm close
in essence to the projection method of Chorin\cite{Chorin1997,Peyret1983})
where one does not need to iterate at each time step. Unfortunately
as said earlier, the strong coupling between energy, density and thermodynamic
pressure needs an iterative scheme and PISO cannot cope with the SCF
system of equations for reasonable time steps without an outer iterative
loop. In our case, the SIMPLE and SIMPLER algorithms have been chosen
to treat the equations and their algorithm is rigorously the same
as the original algorithms described by Patankar\cite{Patankar1980},
the only special treatment being the solution of the density and the
thermodynamic pressure which is shown hereafter.

The van der Waals equation of state being non linear, if written as
$F(\rho)=0$, it has vanishing first derivative $F'(\rho)$ near the
critical point. The latter does not allow for a direct solution using
methods such as Newton-Raphson. So we have to resort to a linearization
of the equation of sate and solve the density iteratively. This done
through the following linearization:

\begin{equation}
\rho^{k+1}=f(\rho^{k})\label{eq13}\end{equation}
 With \begin{equation}
F(\rho)=\frac{\left(P_{0}+a\rho^{2}\right)\left(1-b\rho\right)}{T}\label{eq14}\end{equation}
 or \begin{equation}
F(\rho)=\frac{\left(P_{0}+a\rho^{2}\right)}{T+b\left(P_{0}+a\rho^{2}\right)}\label{eq15}\end{equation}

The convergence rate of such linearization can be found in Accary
\& Raspo \cite{Accary2006}.

In order to close the system, we need an equation to derive the thermodynamic
pressure P$_{0}$. This equation is obtained through the conservation
of mass as follow: \begin{equation}
\int\limits _{\Omega}\rho d\Omega=\int\limits _{\Omega}\rho_{0}d\Omega\label{eq16}\end{equation}
 $\Omega$ being the fluid domain and $\rho_{0}$ the initial density.
It has been found that internal iterations on equation (\ref{eq14})
or (\ref{eq15}) inside the SIMPLER iteration improve drastically
the convergence stability and enables to take bigger time steps than
in the case without inner iterations. When solving $P_{0}$ from equation
(\ref{eq16}) and using Eq. (\ref{eq15}), $P_{0}$ in the numerator
is at iteration k+1 and $P_{0}$ in the denominator at iteration k.

We can summarize the different steps of the resolution as follow for
each time step:

\begin{enumerate}
\item Solve the density field and the thermodynamic pressure using a known
temperature 
\item Solve the momentum equations applying SIMPLER or SIMPLE algorithm
\item Solve conservation of mass 
\item Construct the divergence term and the viscous dissipation term going
to the energy equation through the velocity just calculated at step
3 
\item Solve the energy equation with the previously computed thermodynamic
pressure at step 1 
\item Repeat to step 1 or achieve convergence on a convergence criteria
calculated for each dependent variable 
\end{enumerate}
The slowness of the convergence of the triplet $(T,\ P_{0},\ \rho)$
impacts also on the momentum and mass conservation equations. It is
to speed up consequently the convergence that we have adopted a new
algorithm which is described here after.

\subsection{Description of the improved algorithm 2}

We will solve first the temperature equation (\ref{eq11}) coupled
with the van der Waals equation of state (\ref{eq12}) and the divergence
of the velocity. In order to obtain the divergence of the velocity,
we will make use of the following:

\begin{equation}
\nabla.\vec{V}=\frac{\left(1-b\rho\right)\frac{dP_{0}}{dt}-\nabla.\left(\lambda\nabla T\right)-\phi}{-\left(P_{0}+a\rho^{2}\right)+2a\rho^{2}\left(1-b\rho\right)}\label{eq17}\end{equation}
 The above relation is obtained by taking the material derivative
of the equation of state, and then substituting the appropriate terms
from the mass and energy conservation equations, more details can
be found later in the text equations (\ref{eq:20}-\ref{eq24}). Such
a formulation of the divergence has been used to solve for low-frequency
vibrations in a near critical fluid and transform the term source
in the energy equation without however changing the algorithm of resolution
\cite{Jounet2000,Jounet2002}.

We can notice that all the terms in Eq. (\ref{eq17}) can be calculated
using only $P_{0}$, $\rho$, and T at the exception of the viscous
dissipation term $\Phi$. In the low Mach number approximation the
viscous term is of second order in terms of Mach number and can be
neglected in our case, but for sake of generality, we will leave it
in the equations. As described in the previous algorithm 1, the thermodynamic
pressure $P_{0}$ can be obtained through the conservation of mass
Eq. (\ref{eq16}), the density through the van der Waals equation
of state Eq. (\ref{eq12}). If we want to have an efficient algorithm
we have to decouple the momentum and mass conservation equations from
the state and energy equation. In order to do that we will resort
to an Adams-Bashforth scheme to linearize the $\left(\rho\vec{V}\right)^{n+1}$
term in the - convective term and the dissipation term in the energy
equation as follow: \begin{equation}
\left(\rho\vec{V}\right)^{n+1}=\frac{3}{2}\left(\rho\vec{V}\right)^{n}-\frac{1}{2}\left(\rho\vec{V}\right)^{n-1}\label{eq18}\end{equation}
 \begin{equation}
\phi^{n+1}=\frac{3}{2}\phi^{n}-\frac{1}{2}\phi^{n-1}\label{eq19}\end{equation}
 By linearizing only the $\left(\rho\vec{V}\right)^{n+1}$ instead
of the full convective term including temperature we do not have to
change the schemes previously used for the convective part (hybrid,
power law, quick, smart etc...). The latter is specially useful for
finite volume methods. Whereas in the spectral methods we can use
the Adams-Bashforth discretisation for the full advective-convective
term, the stability of such a scheme is discussed in Ouazzani \& al\cite{Ouazzani1986}.

We have now decoupled at each time step the momentum and mass conservation
equations from the energy and state equations.

It means that through step 1 to 6 at each time step the temperature
can be solved independently of velocities at time n+1 but rather by
using velocities at time n, n-1.

The new algorithm will be as follow for each time step:

\begin{enumerate}
\item Solve the density field and the thermodynamic pressure using a known
temperature 
\item Construct the divergence term using equation (\ref{eq17}) and the
viscous dissipation term described above. 
\item Solve the energy equation with the previously computed thermodynamic
pressure at step 1, and use equation (\ref{eq18}) for velocities
in the convective term. 
\item Repeat to step 1 until convergence is achieved on temperature, thermodynamic
pressure and density 
\item if convergence is obtained go to step 6 
\item Solve the Navier-Stokes equations applying SIMPLER or SIMPLE algorithm
(as shown in figure 1)
\end{enumerate}
\begin{figure}
\includegraphics[width=1\columnwidth]{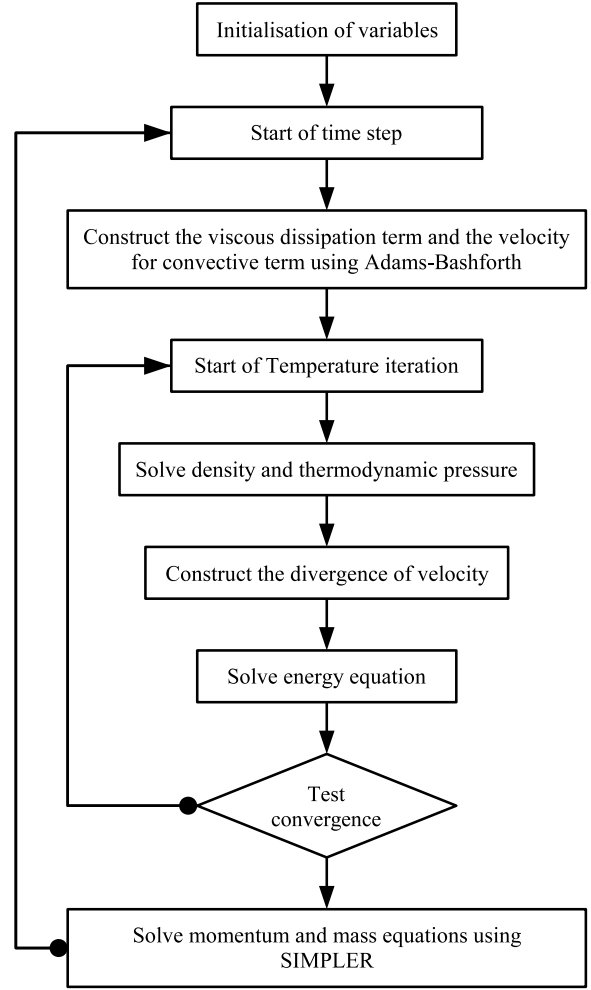} \label{orga}

\caption{Algorithm 2 organigram}
\end{figure}

The convergence speed of these equations becomes similar to the incompressible
equivalent.

By doing in such a manner, we have a better accuracy in solving temperature,
thermodynamic pressure and density; the piston effect can be shown
without having to solve the momentum equation for time of the order
of piston time. This approach enables us after optimization to solve
more easily three dimensional problems.

In a future paper, we will present a spectral method resolution of
SCF and an extension to 3D cylindrical problems.

This treatment can be applied similarly with other equations of state
as follow:

Let's assume that we have the following relation for the EOS:

\begin{equation}
F(\rho,P,T)=0\label{eq:20}\end{equation}

We can then derive this equation and obtain:

\begin{equation}
\frac{d\rho}{dt}\frac{\partial F}{\partial\rho}+\frac{dT}{dt}\frac{\partial F}{\partial T}+\frac{dP}{dt}\frac{\partial F}{\partial P}=0\label{eq21}\end{equation}
 Then using the following two equations in equation (\ref{eq21}):
\begin{equation}
\frac{dT}{dt}=-\frac{T}{\rho}\left(\frac{\partial P}{\partial T}\right)_{\rho}\nabla.\vec{V}+\frac{1}{\rho}\nabla.\left(\lambda\nabla T\right)+\frac{\phi}{\rho}\label{eq22}\end{equation}

\begin{equation}
\frac{d\rho}{dt}=-\rho\nabla.\vec{V}\label{eq23}\end{equation}
 We obtain an equation for the divergence of velocity:

\begin{equation}
\nabla.\vec{V}=\frac{\rho\frac{dP}{dt}\frac{\partial F}{\partial P}+\frac{\partial F}{\partial T}\left(\nabla.\left(\lambda\nabla T\right)+\phi\right)}{\rho\left(\rho\frac{\partial F}{\partial\rho}+\frac{T}{\rho}\left(\frac{\partial P}{\partial T}\right)_{\rho}\frac{\partial F}{\partial T}\right)}\label{eq24}\end{equation}

It follows also if another EOS than van der Waals is chosen that the
other physical transport properties will have also to be redefined
accordingly.

\section{Numerical validation}

For numerical validation purpose, we have chosen two already existing
cases which have been solved extensively by many authors. These two
test cases are the adiabatic heated cavity from one side \cite{Zappoli1996}and
the 2D Rayleigh-Benard problem in a squared cavity\cite{Amiroudine2001,Accary2006,Polezhaev2005}.

\subsection{The 2D adiabatic heated square cavity }

We consider the problem of the interaction between gravitational convection
and the piston effect in a square cavity of 1cm of side dimension
filled with pure CO$_{\text{2}}$ set at 1K above the critical temperature.
All its boundaries are thermally insulated except the one located
at x=0 where the temperature is increased linearly of 10 mK over a
period of 1s. The results obtained with the two methods are very close
(less than 0.1\%) and compare also to the ones obtained by Zappoli
\& al \cite{Zappoli1996}. 

The new algorithm has proved to be more efficient in terms of CPU
time and a factor of 4 has been observed in the case of a mesh of
80 x 80. This speed up is also due to the fact that the resolution
of the energy equation does not need to recompute at each iteration
the {}``$a_{n}$'' coefficients appearing in the finite volume discretisation
of the differential equations. The iterative set coupling temperature,
density and pressure can even be improved by a Newton-Raphson type
method. The optimization of this algorithm will be presented in a
future work with a spectral code.

In Fig 2, we present temperature field obtained at 4.5s. We will not
discuss the physics related to the test cases due to the fact that
they have already been discussed thoroughly in \cite{Zappoli1996}.
We can just point out that we observe a hot spot at the left corner
of the cavity after the 1s heating period and then it is convected
for higher times which confirms the previous results obtained in \cite{Zappoli1996}
and by our work with algorithm 1 presented above.

Through the examination of the steps in the algorithm, we can see
that the piston effect is an adiabatic effect related to a thermodynamic
relation between the heat supplied to the system and the generation
of a pressure work resulting in a fluid motion. In the new algorithm,
the effect is fully embodied in the energy equation coupled with an
appropriate EOS and the conservation of mass at each time step.

\begin{figure}[H]
\includegraphics[width=1\columnwidth,keepaspectratio]{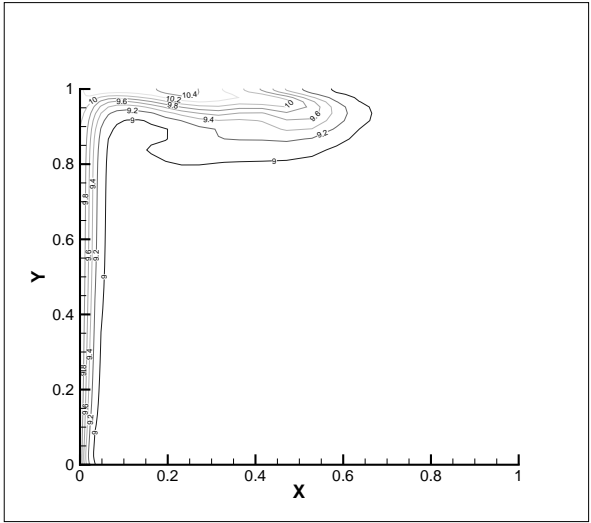} \label{IsoTemp}

\caption{Isotemperature at time t=4.5s with algorithm 2. Distance to the critical
point 1K.}
\end{figure}

\subsection{The 2D Rayleigh Benard problem in a square cavity }

In this second test case, we consider the Rayleigh-Benard \cite{Amiroudine2001,Accary2006,Polezhaev2005}
problem which consists in heating from bottom a square cavity filled
with a supercritical fluid. The fluid is maintained at a constant
temperature in the upper boundary of the cavity whereas the side boundaries
are adiabatic and impermeable. 

The square cavity is a two dimensional cavity with 10mm sides filled
with CO$_{\text{2}}$ on the critical isochore, initially at 1K above
the critical temperature. A mesh of 70 x 90 is considered with strong
refinement of the mesh at the four walls. The time step is kept constant
at 0.05.

The results in this case agree well with those obtained by Amiroudine
\& al\cite{Amiroudine2001}. In figs 3 \& 4, we show the temperature
contours for an increase of 10mK of the bottom wall, for different
times at 6.4s and 8.5s, respectively. We observe the thermal plumes
which appear at the hot and cold walls. The piston effect generates
a cold boundary layer at the top of the cavity where gravitational
instabilities develop as well as at the heated bottom wall. In figure
5, we can see the corresponding velocity field at time 8.5s. 

In this present case, we have the same speed up as for the side heated
2D cavity presented above but furthermore we can increase the time
step by a factor of 10. which leads to a factor of 20 and more of
CPU time gain.

\begin{figure}[H]
\includegraphics[width=1\columnwidth]{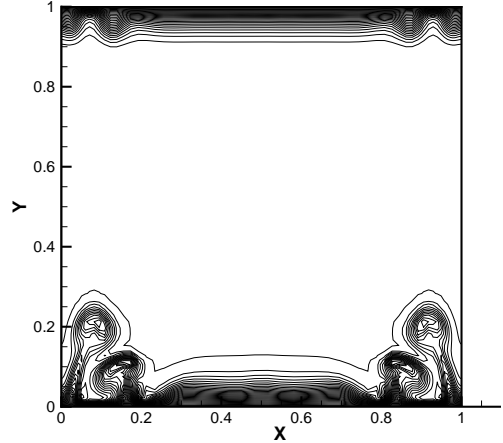} \label{IsoTemp RB1}

\caption{Isotemperature at time t=6.4s with algorithm 2. Distance to the critical
point 1K. Same results as with algorithm 1 with a timestep 10 times
bigger.}
\end{figure}

\begin{figure}[H]
\includegraphics[width=1\columnwidth]{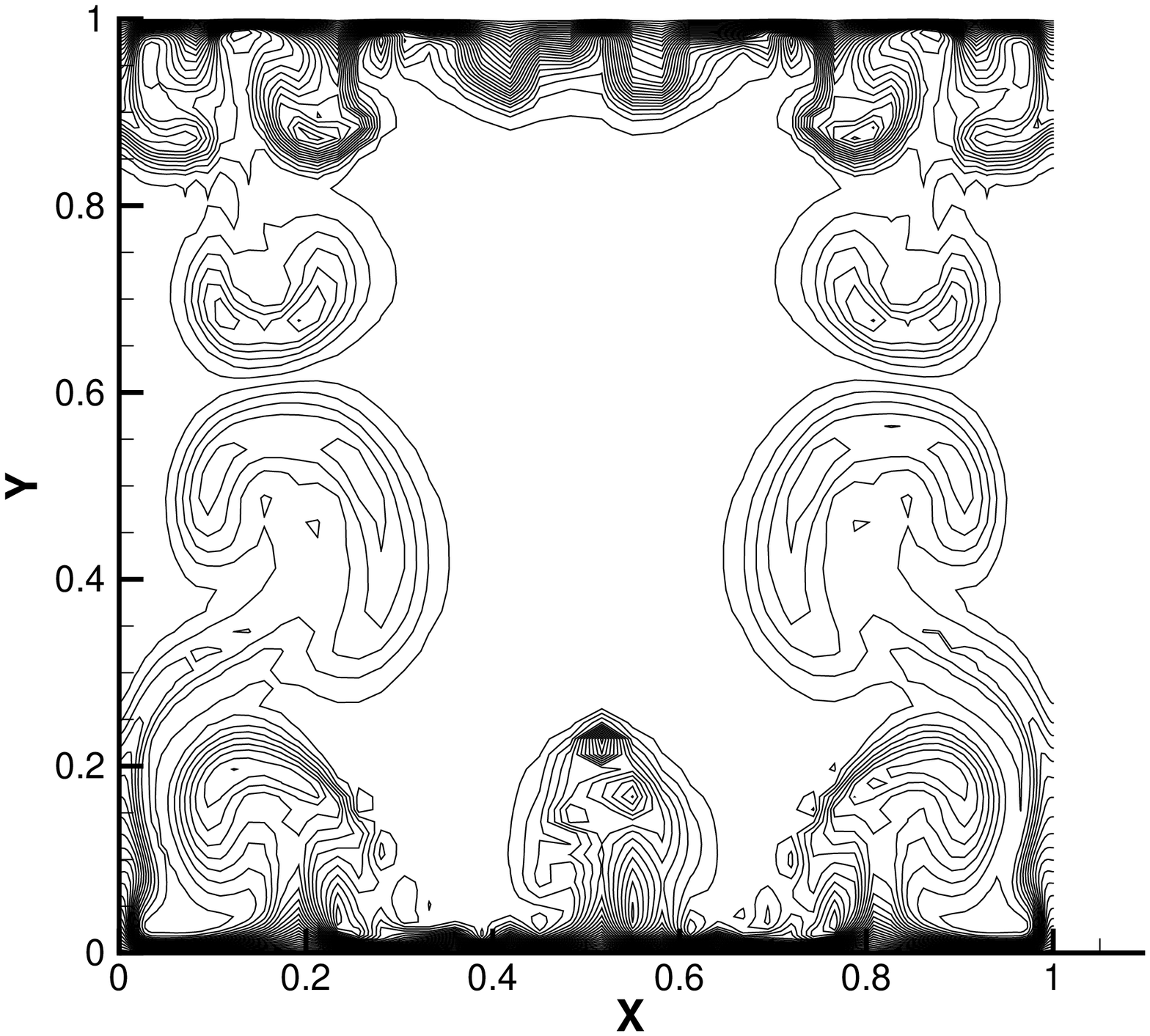} \label{IsoTemp RB2}

\caption{Isotemperature at time t=8.5s with algorithm 2. Distance to the critical
point 1K.}
\end{figure}

\begin{figure}[H]
\includegraphics[width=1\columnwidth]{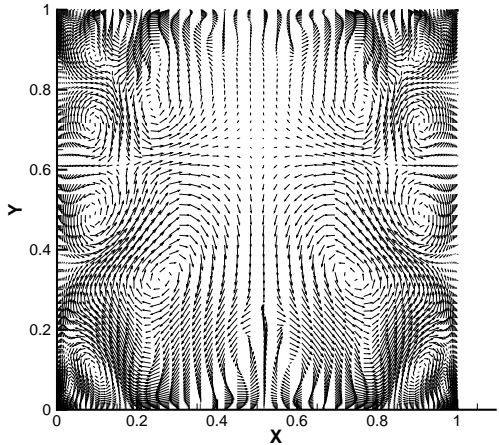} \label{Vec RB2}

\caption{Velocity Field at time t=6s with algorithm 2. Distance to the critical
point 1K.}
\end{figure}

\section{Conclusion}

In this paper, we have developed a new algorithm to solve low Mach
number flows. The algorithm takes advantage of the low Mach number
filtering by computing the divergence of velocity from the three combined
equations (equation of state, mass conservation and energy), by doing
so and using an Adams-Bashforth discretisation for the convective
terms as well as for the viscous dissipation term, one can decouple
at each time step the energy equation from the momentum and mass conservation
equations. This decoupling results in a better stability of the full
algorithm and allows for a substantial saving of CPU time. Even if
time step limitations are introduced through the CFL condition, these
limitations do not lower the time step as compared to the fully implicit
case which is in fact limited in the time step size by physical aspects.

The other benefits of such an algorithm are to be able to use non
iterative methods such as PISO, as well as other discretisation techniques
such as pseudo spectral techniques, and it is easily applied to low
mach number flows in general (as encountered in combustion problems).
Existing code for ideal gas law can be easily modified to tackle supercritical
fluids and 3D problems can be tackled in a reasonable amount of CPU
times. 

One other main reason to introduce such an algorithm for low Mach
number critical fluids is to be able to treat problem with real equations
of state other than van der Waals. These EOS are CPU time consuming
and were not often used whereas now, we are introducing them in our
2D and 3D finite volume code. 

\begin{ack}
The authors would like to thank the CNES for financial support through
the DECLIC program. They thank Fabien Palencia and Eric Georgin for
technical help, and they gratefully thank Bernard Zappoli and Isabelle
Raspo for their remarks and direct help in the writing of this article.
\end{ack}

\end{document}